**Enhanced Prognosis for Abiotic Natural Gas and Petroleum Resources**

J. Marvin Herndon
Transdyne Corporation
San Diego, CA 92131 USA

March 26, 2006

## Abstract

The prognosis for vast potential resources of abiotic natural gas and petroleum depends critically upon the nature and circumstances of Earth formation. Until recently, that prognosis has been considered solely within the framework of the so-called "standard model of solar system formation", which is incorrect and leads to the contradiction of terrestrial planets having insufficiently massive cores. By contrast, that prognosis is considerably enhanced (*i*) by the new vision I have disclosed of Earth formation as a Jupiter-like gas giant; (*ii*) by core formation contemporaneous with raining out from within a giant gaseous protoplanet rather than through subsequent whole-Earth re-melting after loss of gases; (*iii*) by the consequences of whole-Earth decompression dynamics, which obviates the unfounded assumption of mantle convection, and; (*iv*) by the process of mantle decompression thermal-tsunami. The latter, in addition to accounting for much of the heat leaving the Earth's surface, for the geothermal gradient observed in the crust, for substantial volcanism, and possibly for earthquake generation as well, also might enhance the prognosis for future abiotic energy supplies by pressurizing and heating the base of the crust, a potential collection point for abiotic mantle methane or other mantle-derived carbon-containing matter.

Nikolai Kudryavtsev[1] originated what has become the modern Russian-Ukrainian theory of abiotic petroleum, which was widely brought to popular attention in the West by Thomas Gold[2,3]. Kudryavtsev[1] argued that no petroleum resembling the composition of natural crude had been made from plant material in the laboratory, cited examples of petroleum being found in crystalline and metamorphic basement formations or in their overlying sediments, and noted instances of large-scale methane liberation associated with volcanic eruptions. Although still a controversial idea, there is increasing public discussion on the subject and on-going experimentation that supports the feasibility of hydrocarbon formation under deep-Earth conditions, even in the absence of primary hydrocarbons[4]. Ultimately, however, the prognosis for vast potential resources of abiotic mantle and deep-crust natural gas and petroleum depends critically upon the nature and circumstances of Earth formation.

Until recently, since the early 1960s, one idea of planetary formation has so dominated the scientific literature as to become known as the so-called "standard model of solar system formation"[5,6]. That model assumed dust would condense from gases of solar composition at pressures of about $10^{-5}$ bar, which would then gather into progressively larger grains, and become rocks, then planetesimals, and ultimately planets. Along the way, the gases would be lost into space, implying Earth formation from completely de-gassed matter. The abiotic theory of natural gas and petroleum developed in the penumbra of the "standard model of solar system formation". For the reasons described below, that model is wrong. The new vision of Solar System formation that I have disclosed[7,8], together with the causally-related Earth-processes that follow there from[9-12], suggest a greatly enhanced prognosis for future abiotic energy supplies.

I have shown from thermodynamic considerations that condensation from a gas of solar composition at pressures of about $10^{-5}$ bar would not lead to minerals characteristic of those of iron-metal-bearing chondrites, but instead would lead to a highly oxidized condensate[13,14]. The so-called "standard model of solar system formation" is wrong, because it would yield terrestrial planets having insufficiently massive cores[8]. Applied to other planetary systems, the "standard model of solar system formation" necessitates the obtuse postulate of planet migration to explain the observed close-to-star gas giants.

I have shown the consistency of Arnold Eucken's 1944 concept[15] of planets raining out in the central regions of hot, gaseous protoplanets, which would lead to sufficiently reduced condensate to account for the massive cores of the terrestrial planets[8,16]. Moreover, in the Eucken concept, core formation is contemporaneous with condensation and occurs in the presence of primordial gases. The "standard model of solar system formation", on the other hand, presumes planetary melting associated with core formation *after* loss of primordial gases.

Planets generally consist of concentric shells of matter, but there has been no adequate geophysical explanation to account for the Earth's non-contiguous crustal continental rock layer, except by assuming that the Earth in the distant past was smaller and subsequently expanded[17]. The Earth, together with primordial gases, amounts to about 300 Earth-masses, and would comprise a mass similar to Jupiter. That great overburden, I have shown, would lead to the rock-plus-alloy kernel being compressed to about 64% of present diameter, the precise amount required for an initially closed, contiguous continental shell[7,9].

My idea of Earth having initially formed as a Jupiter-like planet is consistent with observations of close-to-star gas giants in other planetary systems. The important point here is that the evidence points to the entirety of Earth formation having taken place in intimate association with primordial gases, which includes about 1.3 Earth-masses of methane. The possibility of carbon-compound occlusion under these conditions is greatly enhanced, relative to the previous Earth-formation concept, and is decidedly relevant to the enhanced prognosis for a deep-Earth methane reservoir and for a carbon-source for abiotic petroleum.

After being stripped of its massive, Jupiter-like overburden of volatile protoplanetary gases, presumably by the high temperatures and/or by the violent activity, such as T Tauri-phase solar wind[18,19], associated with the thermonuclear ignition of the Sun, the Earth would inevitably begin to decompress. The initial whole-Earth decompression is expected to result in a global system of major *primary* cracks appearing in the rigid crust which persist and are identified as the global, mid-oceanic ridge system, just as explained by Earth expansion theory[20]. But here the similarity with that theory ends. In *whole-Earth decompression dynamics*[10], I set forth a different mechanism for global geodynamics which involves the formation of *secondary* decompression cracks and the in-filling of those cracks, a process which is not limited to the last 200 million years, the maximum age of the seafloor.

As the Earth subsequently decompresses and swells from within, the deep interior shells may be expected to adjust to changes in radius and curvature by plastic deformation. As the Earth decompresses, the area of the Earth's rigid surface increases by the formation of secondary decompression cracks often located near the continental margins and presently identified as submarine trenches. These secondary decompression cracks are subsequently in-filled with basalt, extruded from the mid-oceanic ridges, which traverses the ocean floor by gravitational creep, ultimately plunging into secondary decompression cracks, thus emulating subduction, but without mantle convection as previously assumed necessary for plate tectonics theory.

The absence of mantle convection, even with the limited buoyancy-driven mass transport associated with hot-spots, means that, in contrast to previous thinking, there is no reason to assume that the Earth's mantle is significantly de-gassed. One might expect that such a circumstance would greatly enhance the prognosis for future abiotic energy supplies, at least by comparison to previous ideas.

One of the consequences of Earth formation as a giant, gaseous, Jupiter-like planet[7,8], as described by whole-Earth decompression dynamics[9-11], is the existence of a vast reservoir of energy, the stored energy of protoplanetary compression, available for driving geodynamic processes related to whole-Earth decompression. Some of that energy, I have suggested[7,12], is emplaced as heat at the mantle-crust-interface at the base of the crust through the process of *mantle decompression thermal-tsunami*.

As the Earth decompresses, heat must be supplied to replace the lost heat of protoplanetary compression. Otherwise, decompression would lower the temperature, which would impede the decompression process. Heat generated within the core from actinide decay and/or fission[21] or from within the mantle may enhance mantle decompression by replacing the lost heat of protoplanetary compression. The resulting mantle decompression will tend to propagate throughout the mantle, like a tsunami, until it reaches the impediment posed by the base of the crust. There, crustal rigidity opposes continued decompression, pressure builds and compresses matter at the mantle-crust-interface, resulting in compression heating. Ultimately, pressure is released at the surface through volcanism and through secondary decompression crack formation and/or enlargement.

The process of *mantle decompression thermal-tsunami* may account for much of the heat leaving the Earth's surface[22], for the geothermal gradient observed in the crust, for substantial volcanism, and perhaps for earthquake generation as well. That process also might greatly enhance the prognosis for future abiotic energy supplies by pressurizing and heating the base of the crust, a potential collection point for abiotic mantle methane or other mantle-derived carbonaceous matter.

Generally in science, whenever new advances are made, old ideas should be re-examined in light of those advances. In the case of the abiotic origin of natural gas and petroleum, that is especially true, as the advances I have made pertaining to the processes operant during the formation of the Solar System, and to the composition and dynamics of planet Earth, all appear to greatly enhance the prognosis for those abiotic resources.